\documentclass[graybox]{svmult}

\usepackage{mathptmx}       
\usepackage{helvet}         
\usepackage{courier}        
\usepackage{type1cm}        
\usepackage{makeidx}         
\usepackage{graphicx,natbib}        
\usepackage{multicol}        
\usepackage[bottom]{footmisc}
\usepackage{amsmath}
\renewcommand{\th}{{\theta}}

\newcommand{\calD}{{\mathcal D}}
\newcommand{\calN}{{\mathcal N}}
\newcommand{\bfX}{{\mathbf X}}
\newcommand{\hsp}{{\hspace{2ex}}}
\newcommand{\by}{{\mathbf y}}

\def\RR{{\bf R}}

\newcommand{\omegat}{{\omega^{(t)}}}

\makeindex             

\begin{document}

\title*{Checking Simplifying Assumption in bivariate conditional copulas}
\author{Evgeny Levi and Radu V. Craiu}
\institute{Evgeny Levi \at University of Toronto, Department of Statistical Sciences, \email{evgeny@utstat.utoronto.ca}
\and Radu V. Craiu \at University of Toronto, Department of Statistical Sciences, \email{craiu@utstat.toronto.edu}}
%
%
\maketitle

\abstract{The paper considers the problem of establishing data support for the simplifying assumption (SA) in a bivariate conditional copula model. It is known that SA  greatly simplifies the inference for a conditional copula model, but standard tools and methods for testing SA tend to not provide reliable results. 
After splitting the observed data into training and test sets, the method proposed will use a flexible training data Bayesian fit to define tests based on randomization and standard asymptotic theory.  
Theoretical justification for the method is provided and its performance is studied   using simulated data.  The paper also discusses implementations  in alternative models of interest, e.g. Gaussian, Logistic and Quantile regressions.}

\section{Introduction}   
\label{sec:intro}
A copula is   mathematical concept which is often used to model the joint distribution of  several random variables.  The applications of copula models permeate a number of fields where of interest is also the dependence structure between the random variables considered, e.g. \cite{Hougaard:2000}, \cite{patton2006modelling}, \cite{Dupuis:2007qo},\cite{Genest:2007fh} and \cite{Lakhal:2008sh}. The propagation of copula-related ideas in probability and statistics started with \cite{Sklar:1959} which proved that for a random vector $(Y_1,\ldots,Y_p)$ with cumulative distribution function (CDF) $H(y_1,\ldots,y_p)$ and marginal continuous CDFs $F_i(y_i)$, $i=1,\ldots,p$  there exists a unique \textit{copula} $C:[0,1]^p\rightarrow [0,1]$ such that
\begin{equation}
H(y_1,\ldots,y_k)=C(F_1(y_1),\ldots,F_k(y_k)).
\end{equation}
For statistical modelling it is also useful to note that,   a p-dimensional copula $C$ and marginal continuous CDFs $F_i(y_i)$, $i=1,\ldots,p$ are  building blocks for   a valid $p$-dimensional CDF,  $C(F_1(y_1),\ldots,F_p(y_p))$  with $i$th marginal CDF equal to $F_i(y_i)$, thus providing much-needed flexibility in modelling multivariate distributions. \\
The above results can be extended when conditioning on a covariate vector $X\in \RR^q$ \citep{lamb-van, patton2006modelling} so that 
\begin{equation}
H(y_1,\ldots,y_k|X)=C_X(F_1(y_1|X),\ldots,F_k(y_k|X)),
\label{cc}
\end{equation}
where all CDFs and the copula are conditional on  $X$. For the rest of this paper we follow \cite{levi2018bayesian} and assume that the copula in \eqref{cc} belongs to a parametric family and its one-dimensional parameter  depends on $X$ through some unknown function $\th(X):\RR^q \rightarrow \Theta$. The range of $\th(X)$ is usually restricted, so we introduce a known one-to-one link function $g:\Theta \rightarrow \RR$ such that the \textit{calibration function}, $\eta:\RR^q \rightarrow \RR$, defined as $\eta(X)=g(\th(X))$ has unrestricted range.\\ 
The simplifying assumption (SA) \citep{czado2010pair}
states that $\eta(X)$ is constant. Clearly, SA  greatly simplifies the estimation in conditional copula models, including their use in hierarchical models such as vines  \citep[see, for instance,][]{czado2009}. It has been shown in \cite{levi2018bayesian} that SA is violated  when important covariates are not included in the model \eqref{cc}. In an important contribution,  \cite{agn} showed that  assuming SA when the data generative process has non-constant calibration may lead to biased results. 
 In light of these results,  there is a genuine demand for  strategies that effectively test whether the SA is appropriate or not. A number of research contributions address this issue for frequentist analyses, e.g. \cite{acar2013statistical}, \cite{gijbels2015estimation}, \cite{derumigny2016tests}\cite{killiches2017examination}.   \\
We place the problem in a Bayesian analysis context where inference for $\eta$ relies on a flexible model, following the general philosophy expounded in \cite{sabeti2014additive, klein},\cite{hernandez2013gaussian} or \cite{levi2018bayesian}.  Within the  Bayesian paradigm, it was observed in \cite{cra-sabeti} that when generic model selection criteria to identify data support for SA tend to favour the more complex model even when SA holds. 
In the next section we present the problem in mathematical terms and review some of the Bayesian model selection procedures one can use in this context. A new approach for testing  SA, based on a data-splitting procedure,  is described in  Section~\ref{sec:meth}.  A merit of the proposal is that it is quite general in its applicability, but this comes, unsurprisingly, at the expense of power. In order to investigate whether the trade-off is reasonable we design a simulation study and present its conclusions in Section~\ref{sec:sim}.  Section~\ref{sec:theory} contains theoretical justification of  the proposed algorithm and the paper closes with a discussion of  extensions to other regression problems and concluding remarks.
\section{The Problem}
\label{sec:prob}
Here we focus on bivariate response variables so that the observed data  consist of $n$ independent triplets $\calD=\{(x_i,y_{1i},y_{2i}),\hsp i=1,\ldots, n\}$ where $y_{1i}$ and $y_{2i}$ are in $\RR$ and $x_i\in \RR^q$. Also let us denote $\by_1=(y_{11},\ldots,y_{1n})$, $\by_2=(y_{21},\ldots,y_{2n})$ and $\bfX \in \RR^{n\times q}$  is the matrix with $i^{th}$ row equal to $x_i^T$. We rely on \eqref{cc} to express the \textit{full conditional model} for  $Y_1$ and $Y_2$ given $X$
\begin{equation}
\label{eq:gen}
P(\omega|\by1,\by2,\bfX)=\prod_{i=1}^n f_1(y_{1i}|\omega_1,x_i)f_2(y_{2i}|\omega_2,x_i)c_{\th(x_i)}\left(F_1(y_{1i}|\omega_1,x_i),F_2(y_{2i}|\omega_2,x_i)\right),
\end{equation}
where $f_j$, $F_j$ are the density and, respectively, the CDF for $Y_j$,  while $\omega_j$  denotes  all the latent variables and parameters associated with the $j$th marginal distribution, for $j=1,2$. The copula density function is denoted by $c$ and it depends on $X$ through unknown function $\th(X)=g^{-1}(\eta(X))$. Note that the expression above is very general with no assumptions about marginal distributions. Assuming that all parameters $\omega$ can be estimated, the  copula family can be selected using several  model selection criteria  \citep[e.g.,][]{sabeti2014additive, levi2018bayesian}. Once the copula family is selected, the objective is to check whether the SA is valid, in other words whether full model in \eqref{eq:gen} becomes the \textit{reduced model}
\begin{equation}
P(\omega|\by1,\by2,\bfX)=\prod_{i=1}^n f_1(y_{1i}|\omega_1,x_i)f_2(y_{2i}|\omega_2,x_i)c_{\th}\left(F_1(y_{1i}|\omega_1,x_i),F_2(y_{2i}|\omega_2,x_i)\right).
\label{cc-sa}
\end{equation}
Note that in \eqref{cc-sa} the copula  depends only on one scalar parameter,  $\th$. \\
If  flexible models as Gaussian Processes are implemented within the Bayesian paradigm, then the characteristics of the posterior distribution will be estimated using  draws $\{\omegat\}_{t=1}^M$ obtained by running an Markov chain Monte Carlo (MCMC) algorithm (e.g.,\cite{sabeti2014additive, levi2018bayesian}) to sample the posterior. Data support  for the full and reduced models, \eqref{eq:gen} and \eqref{cc-sa}, may be  established using several  criteria. We briefly review two options that distinguish between models based on predictive power.\\
\subsubsection*{The Cross-Validated Pseudo Marginal Likelihood and Its Conditional Variant}
The cross-validated pseudo marginal likelihood (CVML) \cite{geisser1979predictive} calculates the average (over parameter values) prediction power for model $\mathcal M$ via
\begin{equation}
\mbox{CVML}(\mathcal M)=\sum_{i=1}^{n}\log\left(P(y_{1i},y_{2i}|\mathcal D_{-i},\mathcal M)\right),
\label{cvml-def}
\end{equation}
where $\mathcal D_{-i}$ is the data set from which the $i$th observation has been removed.
An estimate of \eqref{cvml-def} for a given model is estimated using posterior draws $\omegat$ given the whole data set $\calD$  \citep[detailed derivations can be found in][]{levi2018bayesian} via
\begin{equation}
\mbox{CVML}_{est}(\mathcal M)=-\sum_{i=1}^{n}\log\left(\frac{1}{M}
\sum_{t=1}^{M}P(y_{1i},y_{2i}|\mathbf \omega^{(t)},\mathcal M)^{-1}\right).
\end{equation}
The model with the largest CVML is preferred. 

The conditional CVML (CCVML), introduced by \cite{levi2018bayesian} specifically for selection of copula models, considers conditional  rather than joint predictions 
\begin{equation}
\mbox{CCVML}(\mathcal M)=\frac{1}{2}\left\{ \sum_{i=1}^{n}\log\left[P(y_{1i}|y_{2i},\mathcal D_{-i},\mathcal M)\right] + \sum_{i=1}^{n}\log\left[P(y_{2i}|y_{1i},\mathcal D_{-i},\mathcal M)\right] \right\}.
\end{equation} 
Again this criterion can be estimated from posterior samples using
\begin{eqnarray}
\mbox{CCVML}_{est}(\mathcal M)&=&-\frac{1}{2}\sum_{i=1}^{n}\left\{\log\left[\frac{1}{M}
\sum_{t=1}^{M}\frac{P(y_{2i}|\mathbf \omega^{(t)},\mathcal M)}{P(y_{1i},y_{2i}|\mathbf \omega^{(t)},\mathcal M)}\right] \right. \nonumber \\
&+&\left. \log\left[\frac{1}{M}
\sum_{t=1}^{M}\frac{P(y_{1i}|\mathbf \omega^{(t)},\mathcal M)}{P(y_{1i},y_{2i}|\mathbf \omega^{(t)},\mathcal M)}\right] \right\}.
\end{eqnarray}
Similar to CVML, the model with the largest CCVML is selected.
\subsubsection*{Watanabe-Akaike Information Criterion}
The Watanabe-Akaike Information Criterion \citep{watanabe2010asymptotic} is an information-based criterion that is closely related to  CVML, as discussed in \cite{vehtari2017practical}.The WAIC is defined as 
\begin{equation}
\mbox{WAIC}(\mathcal M)=-2\mbox{fit}(\mathcal M) + 2\mbox{p}(\mathcal M) ,
\label{waic}
\end{equation}
where the model fitness is 
\begin{equation}
\mbox{fit}(\mathcal M)=\sum_{i=1}^{n}\log E\left[P(y_{1i},y_{2i}|\mathbf \omega,\mathcal M)\right]
\label{eq:fitness}
\end{equation}
and the penalty 
\begin{equation}
\mbox{p}(\mathcal M)=\sum_{i=1}^{n}\mbox{Var}[ \log P(y_{1i},y_{2i}|\mathbf \omega,\mathcal M)]. 
\label{eq:penalty}
\end{equation}
The expectation in \eqref{eq:fitness} and the variance in \eqref{eq:penalty} are with respect to the conditional distribution of $\omega$ given the data and can easily be estimated using the $\omegat$ draws.
The model with the smallest WAIC measure is preferred.
\section{Detecting Data Support for SA}
\label{sec:meth}
As will be shown in Section~\ref{sec:sim} the criteria described above have unsatisfactory performances when the reduced model is the generative one. Instead, we propose to use some of the  properties  that are invariant to the group of permutations when SA indeed holds. 
In the first stage we randomly divide the data $\calD$ into training and test sets, $\calD_1$ and $\calD_2$,  with $n_1$ and $n_2$ sample sizes, respectively. The full model defined by \eqref{eq:gen} is fitted on $\calD_1 $, and we denote $\omegat$ the $t$-th draw sampled  from the posterior. For the $i$th item in $\calD_2$, compute point estimates $\hat \eta_i$ and $\hat U_i=(\hat U_{1i},\hat U_{2i})$ where $\hat U_{ji} = F_j(y_{ji}|\hat\omega_j,x_i)$, $j=1,2$, $i=1,\ldots,n_2$. The   marginal parameters estimates, $\hat\omega_j$,  are  obtained from the training data posterior draws.  For instance, if the marginal models are $Y_{1i}\sim \calN(f_1(x_i),\sigma_1^2)$ and $Y_{2i}\sim \calN(f_2(x_i),\sigma_2^2)$, then each of the MCMC sample $\omegat$ leads to an estimate $\hat f_1^t(x_i),\hat f_2^t(x_i),\hat \sigma^t_1,\hat \sigma^t_2,\hat \eta^t(x_i)$. 
Then $\hat U_i=(\hat U_{1i},\hat U_{2i})$  are obtained using $$(\hat U_{1i},\hat U_{2i})=(\Phi((y_{1i}-\overline{\hat f_1(x_i)})/\overline{\hat \sigma_1}),\Phi((y_{2i}-\overline{\hat f_2(x_i)})/\overline{\hat \sigma_2})),$$
where the overline $\overline{a}$ signifies the averages of Monte Carlo draws $a^t$.

Given the vector of  calibration function evaluations at the test points, $\hat \eta=(\hat \eta_1,\ldots,\hat \eta_{n_2})$, and a partition $\min(\hat \eta)=a_1<\ldots <a_{K+1}=\max(\hat \eta)$ of the range of $\eta$ into $K$ disjoint intervals, define the set of observations in $\calD_2$ that yield calibration function values between $a_k$ and $a_{k+1}$, $B_k=\{i:a_k \leq \hat \eta_i < a_{k+1}\}$  $k=1,\ldots,K$. We choose the partition such that each "bin" $B_k$ has approximately the same number of elements, $n_2/K$.

\begin{table}[h]
\label{tab:met1}
\begin{enumerate}
\item[{\bf A1}]  Compute the $k$th bin-specific Spearman's rho $\hat \rho_k$ from $\{\hat U_i:i \in B_k\})$ $k=1,\ldots,K$. 
\item[{\bf A2}] Compute the observed statistic $T^{obs}=\max_k(\hat\rho_k)-\min_k(\hat\rho_k)$. Note that if 
      SA holds, we expect the observed statistic to be close to zero.
\item[{\bf A3}] Consider $J$ permutations $\lambda_j:\{1,\ldots,n_2\}\rightarrow\{1,\ldots,n_2\}$. For each permutation $\lambda_j$:
 \begin{enumerate}
    \item[A3.1] Compute $\hat \rho_{jk} = \rho(\{\hat U_i:\lambda_j(i) \in B_k\})$ $k=1,\ldots,K$.
     \item[A3.2] Compute test statistic $T_j=\max_k(\hat\rho_{jk})-\min_k(\hat\rho_{jk})$. Note if SA holds, then we expect $T_j$ to be close to $T^{obs}$.
  \end{enumerate}
\item[{\bf A4}] We consider that there is support in favour of SA at significance level $\alpha$ if $T^{obs}$ is smaller  than the $(1-\alpha)$-th empirical quantile calculated from the sample $\{T_j: 1\le j \le J\}$.
\end{enumerate}
\caption{\it Method 1: A permutation-based procedure for assessing data support in favour of SA}
\end{table}

Under SA, the bin-specific estimates for various measures of dependence, e.g. Kendall's $\tau$ or Spearman's $\rho$, computed from the samples $\hat U_i$, are invariant to permutations, or swaps across bins. Based on this  observation, we consider the   procedure described in Table 1 for identifying data support for SA. The distribution of the resulting test statistics obtained in Method 1 is determined empirically, via permutations. Alternatively, one can rely on the asymptotic properties of the bin-specific dependence parameter estimator and construct Chi-square test. Specifically, suppose  the bin-specific Pearson correlations $\hat \rho_k$  are computed from samples
$\{\hat U_i:i \in B_k\})$, for all $k=1,\ldots,K$. Let $\hat \rho = (\hat \rho_1,\ldots,\hat \rho_K)^T$, and $\tilde n = n_2/K$ be the number of points in each bin. It is known that $\hat\rho_k$ is asymptotically normal distributed for each $k$ so that  $$\sqrt{\tilde n}(\hat \rho_k-\rho_k) \overset{d}{\rightarrow} \calN(0,(1-\rho_k^2)^2),$$ where $\rho_k$ is the true correlation in bin $k$. If we assume that $\{\hat \rho_k$:\; $k=1,\ldots,K\}$ are independent, and set  $\rho = (\rho_1,\ldots,\rho_K)^T$ and $\Sigma=diag((1-\rho_1^2)^2,\ldots,(1-\rho_K^2)^2)$, then we have: 
$$\sqrt{\tilde n}(\hat \rho - \rho)\overset{d}{\rightarrow} \calN(0,\Sigma)$$

In order to combine evidence across bins, we define the matrix $A \in \RR^{(K-1)\times K}$ as
$$A = \begin{bmatrix} 1 & -1 & 0 & \cdots & 0 \\
                    0 &  1 & -1 & \cdots & 0 \\
                    \vdots & \vdots & \vdots & \vdots & \vdots \\
                    0 & 0 & \cdots & 1 & -1 \end{bmatrix}$$
Since under the null hypothesis SA holds, one gets  $\rho_1= \ldots =\rho_K$, implying
$$\tilde n (A\hat\rho)^T (A\Sigma A^t)^{-1} (A\hat \rho) \overset{d}{\rightarrow} \chi^2_{K-1}.$$ Method 2, with its  steps  detailed in  Table \ref{tb:met2}, relies on the ideas above to test SA.

\begin{table}
\begin{enumerate}
\item[{\bf B1}] Compute the bin-specific Pearson correlation $\hat \rho_k$  from samples
$\{\hat U_i:i \in B_k\})$, for all $k=1,\ldots,K$. Let $\hat \rho = (\hat \rho_1,\ldots,\hat \rho_K)^T$, and $\tilde n = n_2/K$, the number of points in each bin.

\item[{\bf B2}] 
Define $\rho = (\rho_1,\ldots,\rho_K)^T$, $\Sigma=diag((1-\rho_1^2)^2,\ldots,(1-\rho_K^2)^2)$ and $A \in \RR^{(K-1)\times K}$ be equal to 
$$A = \begin{bmatrix} 1 & -1 & 0 & \cdots & 0 \\
                    0 &  1 & -1 & \cdots & 0 \\
                    \vdots & \vdots & \vdots & \vdots & \vdots \\
                    0 & 0 & \cdots & 1 & -1 \end{bmatrix}.$$
Compute $T^{obs} = \tilde n (A\hat\rho)^T (A\hat \Sigma A^t)^{-1} (A\hat \rho)$.
\item[{\bf B3}] Compute p-value = $P(\chi^2_{K-1} > T^{obs})$ and reject SA if p-value$<\alpha.$ 
\end{enumerate}
\label{tb:met2}
\caption{\it Method 2:  A Chi-square test for assessing data support in favour of SA}\end{table}

Method 1 evaluates the p-value using a randomization procedure \cite{lehmann2006testing}, while  the second is based on the asymptotic normal theory of Pearson correlations. To get reliable results it is essential to assign test observations to "correct" bins which is true when calibration predictions are as close as possible to the true unknown values, i.e.  $\hat \eta(x_i)\approx \eta(x_i)$. The latter heavily depends on the estimation procedure and sample size of the training set. Therefore it is advisable to apply very flexible methodologies for the calibration function estimation and have enough data points in the training set.  We immediately see a tradeoff as more observations are assigned to $\calD_1$ the better will be the calibration test predictions, at the expense of decreasing power due to a smaller sample size in $\calD_2$. For our simulations we have used $n_1\approx 0.65n$ and $n_2\approx 0.35n$, and $K\in\{2, 3\}$.\\

\section{Simulations}
\label{sec:sim}
In this section we present the performance of the proposed methods and comparisons with generic CVML and WAIC criteria on simulated data sets. Different functional forms of calibration function, sample sizes and magnitude of deviation from SA will be explored.
\subsubsection*{Simulation details}
We generate samples of sizes $n=500$ and $n=2000$ from 6 scenarios described below. For all scenarios Clayton copula \cite{embrechts2001modelling} will be used to model dependence between responses, covariates are independently sampled from $\mathcal U[0,1]$. For scenarios 1 to 3, the covariate dimension $q=2$ in the remaining ones $q=5$. Marginal conditional distributions $Y_1|X$ and $Y_2|X$ are modeled as Gaussian with constant variances $\sigma_1^2,\sigma_2^2$ and conditional means $f_1(X),f_2(X)$ respectively. All these parameters are generally not known in advanced and must be estimated jointly with the calibration function $\eta(X)$. For convenience we parametrize calibration by Kendall's tau $\tau(X)$ \cite{embrechts2001modelling} which has one-to-one correspondence with $\eta(X)$ but takes values in $[-1,1]$.
\begin{enumerate}
\item[\textbf{Sc1}] $f_1(X)=0.6\sin(5x_1)-0.9\sin(2x_2)$,\\
$f_2(X)=0.6\sin(3x_1+5x_2)$,\\
$\tau(X)=0.5$,$\sigma_1=\sigma_2=0.2$ 
\item[\textbf{Sc2}] $f_1(X)=0.6\sin(5x_1)-0.9\sin(2x_2)$,\\
$f_2(X)=0.6\sin(3x_1+5x_2)$,\\
$\tau(X)=0.5+\beta\sin(10X^T\beta)$\\
$\beta=(1,3)^T/\sqrt{10}$, $\sigma_1=\sigma_2=0.2$ 
\item[\textbf{Sc3}] $f_1(X)=0.6\sin(5x_1)-0.9\sin(2x_2)$,\\
$f_2(X)=0.6\sin(3x_1+5x_2)$,\\
$\tau(X)=0.5 + \beta*2*(x_1+\cos(6x_2)-0.45)/3$\\
$\sigma_1=\sigma_2=0.2$ 
\end{enumerate}
\textbf{Sc1} corresponds to SA since Kendall's tau is independent of covariate level. The calibration function in \textbf{Sc2} has single index form for the calibration function, while in \textbf{Sc3}  it has an additive structure on $\tau$ scale (generally not additive on $\eta$ scale), these simulations are useful to evaluate performance under model mispecification. Also tau  in \textbf{Sc2} and \textbf{Sc3} depend on parameter $\beta$, which in this study is set to $\beta=0.25$. 
\subsubsection*{Simulation results}
For each sample size and  scenario we have repeated the analysis using 250 independently replicated data sets. For each data, the GP-SIM model suggested by \cite{levi2018bayesian} is fitted. This method implements sparse Gaussian Process (GP) priors for marginal conditional means and sparse GP-Single Index for calibration function. The inference is  based on 5000 MCMC samples for all scenarios, as the chains were run for 10000 iterations with 5000 samples discarded as burn-in. The number of inducing inputs was set to 30 for all GP. For generic SA testing, GP-SIM fitting is done for the whole data sets and posterior draws are used to estimate CVML and WAIC. Since the proposed methods require splitting the data set into training and test sets, we first randomly divide each data set with proportion 65\% to training and 35\% for testing then fit GP-SIM on training set and then use posterior draws for point estimates of $F_1(y_{1i}|x_i)$, $F_2(y_{2i}|x_i)$ and $\eta(x_i)$ for every observation in test set. In Method 1 we used 500 permutations.  Table~\ref{table:gen} shows the percentage of SA rejections for  $\alpha=0.05$.     
\begin{table} [!ht]
\begin{center}
\caption{ Simulation Results: Generic, proportion of rejection of SA for each scenario, sample size and generic criteria.}
\smallskip
\scalebox{1.2}{
\begin{tabular}{l| l l l || l l l  }
\multicolumn{1}{c}{ }  &  \multicolumn{3}{c}{$\mathbf{N=500}$}   &\multicolumn{3}{c}{$\mathbf{N=1000}$} \\
\hline
Scenario                     & CVML       &  CCVML     &   WAIC      &  CVML       &   CCVML      &  WAIC \\
\hline
\textbf{Sc1}                 & $43.5\%$   & $44.0\%$   &  $43.5\%$   &  $32.9\%$   & $26.2\%$     &  $33.9\%$  \\
\textbf{Sc2}   & $100\%$    & $100\%$    &  $100\%$    &  $100\%$    &  $100\%$     &  $100\%$    \\
\textbf{Sc3}   & $100\%$    & $100\%$    &  $100\%$    &  $100\%$    &  $99.6\%$    &  $100\%$   \\
\hline
\end{tabular}
}
\label{table:gen}
\end{center}
\end{table}
The presented results clearly illustrate that generic methods have very high  type I error probabilities. This leads to a loss of statistical efficiency since a complex model is selected over a much simpler one. In addition, the SA may be of interest in itself in certain applications, e.g. stock exchange modelling where it is useful to determine whether the dependence structure between different stock prices does not depend on other factors.

\begin{table} [!ht]
\begin{center}
\caption{ Simulation Results: Proposed method, proportion of rejection of SA for each scenario, sample size, number of bins (K) and  method.}
\smallskip
\scalebox{1.1}{
\begin{tabular}{l| l l l l  || l l l l  }
\multicolumn{1}{l}{} &  \multicolumn{4}{c}{Permutation test}  &  \multicolumn{4}{c}{$\chi^2$ test}  \\
\hline
&\multicolumn{2}{c}{$\mathbf{N=500}$} &  \multicolumn{2}{c||}{$\mathbf{N=1000}$} & \multicolumn{2}{c}{$\mathbf{N=500}$} &  \multicolumn{2}{c}{$\mathbf{N=1000}$}\\
\hline
Scenario                   &  $K=2$     &   $K=3$    &  $K=2$     &   $K=3$    &   $K=2$    &   $K=3$    &  $K=2$      &   $K=3$      \\
\hline
\textbf{Sc1}               &  $6.7\%$   & $6.2\%$    & $4.9\%$   &  $4.0\%$   & $8.9\%$   &  $9.7\%$   &  $8.0\%$   &  $8.4\%$    \\
\textbf{Sc2}($\beta=0.25$)  &  $97.3\%$  &  $96.4\%$  &  $100\%$   &  $100\%$   &  $100\%$  &  $99.1\%$  &  $100\%$  &  $100\%$   \\
\textbf{Sc3}($\beta=0.25$)  &  $56.4\%$  &  $49.3\%$  &   $88.9\%$  &  $88.4\%$ &  $69.3\%$ &  $68.4\%$  &  $96.8\%$ &  $96.9\%$   \\
\hline
%
\end{tabular}
}
\label{table:bin}
\end{center}
\end{table}

The simulations summarized in Table~\ref{table:bin} show that the proposed methods  have much smaller probability of Type I error which vary around the threshold of 0.05. It must be pointed, however, that under SA the performance of $\chi^2$ test worsens with the number of bins $K$, which is not surprising since as $K$ increases, the number of observations in each bin goes down and normal approximation for the distribution of Pearson correlation becomes tenuous. The performance of both methods  improves with sample size. We also notice a  loss of power between Scenarios 2 and  3, which is due to model misspecification, since in the latter case the generative model is different from the postulated one.

\section{Theoretical Discussion}
\label{sec:theory}
In this section we prove that under canonical assumptions, the probability of type I error for Method 2 in Section 3 converges to $\alpha$ when SA is true. 


Suppose we have independent samples from $K$ populations (groups) with the same sample size, $(u_{1i}^1,u_{2i}^1)_{i=1}^{n_1}\sim (U_1^1,U_2^1)$,\ldots,$(u_{1i}^K,u_{2i}^K)_{i=1}^{n_K}\sim (U_1^K,U_2^K)$, the goal is to test $\rho_1=\ldots=\rho_K$. To simplify notation, we assume $n_1=\ldots, n_K=n$. Letting $\hat \rho = (\hat \rho_1,\ldots,\hat \rho_K)$ be a vector of sample correlations , $\Sigma=diag((1-\rho_1^2)^2,\ldots,(1-\rho_K^2)^2)$ and $(K-1)\times K$ matrix A as defined in Section~\ref{sec:meth}, then canonical asymptotic results imply that, as $n \rightarrow \infty$,
\begin{equation}
\label{eq:chisq-test}
T = n(A\hat\rho)^T(A\Sigma A^T)^{-1}(A\hat \rho) \overset{d}{\rightarrow} \chi^2_{K-1}.
\end{equation}
Based on the model fitted on $\calD_1$, we define estimates of $F_1(y_{1i}|x_i)$ and $F_2(y_{2i}|x_i)$ by $\hat U=\{\hat U_i=(\hat F_1(y_{1i}|x_i),\hat F_1(y_{2i}|x_i))\}_{i=1}^{n_2}$. Note that $\hat U$ depends on $\calD_1$ and $X$. Given a fixed number of bins $K$ and assuming, without loss of generality, equal sample sizes in each bin $\tilde n = n_2/K$, the first step is to assign $\hat U_i$ to bins by values of $\hat \eta(x_i)$. Introduce a permutation $\lambda^*:\{1,\ldots,n_2\}\rightarrow \{1,\ldots,n_2\}$ that "sorts" $\hat U$ from smallest $\hat \eta(x)$ value to largest i.e. $\hat U_{\lambda^*}=\{\hat U_{\lambda^*(i)}\}_{i=1}^{n_2}$ with $\hat \eta(x_{\lambda^*(1)})< \hat \eta(x_{\lambda^*(2)}) <\cdots < \hat \eta(x_{\lambda^*(n_2)})$. Finally define the test function $\phi()$ with specified significance level $\alpha$ to test SA:
\begin{equation}
\phi(\hat U|\calD_1,X,\lambda^*)=\begin{cases} 1 \hsp \mbox{ if } \hsp T(\hat U_{\lambda^*})>\chi^2_{K-1}(1-\alpha) \\ 
                                               0 \hsp \mbox{ if } \hsp T(\hat U_{\lambda^*})\leq\chi^2_{K-1}(1-\alpha). \end{cases}
\end{equation}
Where test function $T(U)$ as in \eqref{eq:chisq-test} with $\hat \rho_1 = \rho(U_1,\ldots,U_{\tilde n})$,$\hat \rho_2 = \rho(U_{\tilde n +1},\ldots,U_{2\tilde n})$,\ldots,$\hat \rho_K = \rho(U_{(K-1)\tilde n+1},\ldots,U_{K\tilde n})$. Intuitively if SA is false then we would expect  $T(\hat U_{\lambda^*})$ to be larger then the critical value $\chi^2_{K-1}(1-\alpha)$. \\
The goal is to show that this procedure have probability of type I error equal to $\alpha$, which is equivalent to the expectation of the test function:
\begin {equation}
\mbox{P(Type I error)} = \int \phi(\hat U|\calD_1,X,\lambda^*)P(\lambda^*|\calD_1,X)P(\hat U|\calD_1,X)P(\calD_1)P(X)d\hat U d\calD_1 dX d \lambda^*.
\end{equation}  
Note that $\lambda^*$ does not depend on $\hat U$ because of the data splitting to train and test sets. Also usually $P(\lambda^*|\calD_1,X)$ is just a point mass at some particular permutation. In general the above integral cannot be evaluated, however if we assume that for all test cases:
\begin{equation}
\begin{split}
&\hat F_1(y_{1i}|x_i) \overset{p}{\rightarrow} F_1(y_{1i}|x_i) \hsp \mbox{as} \hsp n\to \infty , \\
&\hat F_2(y_{2i}|x_i) \overset{p}{\rightarrow} F_2(y_{2i}|x_i) \hsp \mbox{as} \hsp n\to \infty .
\end{split}
\end{equation}
Then under SA and as $n\to \infty$, $P(\hat U|\calD_1,X)\approx \prod_{i=1}^{n_2}c(\hat u_{1i},\hat u_{2i})$ where $c(,)$ is copula density and the expectation becomes:
\begin {equation}
\begin{split}
\mbox{P(Type I error)} =& \int \phi(\hat U|\lambda^*)P(\lambda^*|\calD_1,X)P(\hat U)P(\calD_1)P(X)d\hat U d\calD_1 dX,d \lambda^* = \\
&\int \left(\int \phi(\hat U|\lambda^*)P(\hat U)d\hat U \right)P(\lambda^*|\calD_1,X)P(\calD_1)P(X)d\calD_1 dX d \lambda^*=\alpha.
\end{split}
\end{equation}  
Therefore if marginal CDF predictions for test cases are consistent then this procedure has the required probability of type I error for sufficiently large sample size. 

\section{Conclusion}
\label{sec:conc}
In this paper we propose two methods to check data support for the  simplifying assumption in conditional bivariate copula problems. The method is based on splitting the whole data set to train and test sets, then partitioning test set into bins using predicted calibration values and finally use randomization or $\chi^2$ test to check if the distribution in each bin is the same or not. It was presented theoretically and empirically that under SA probability of Type I error is controlled while generic methods fail to provide reliable results. In addition to conditional copulas we also mentioned how this idea can be generalized to variate of different problems. There are still some uncertainty about what proportion of the data should be assigned to train and which to test set. It was also assumed that sample sizes in each "bin" is the same however in some problems power can be increases by changing sample sizes in each bin. These problems will be investigated further.

\bibliographystyle{ims}
\bibliography{Ref}

\end{document}